\documentclass[smallextended]{svjour3}       % onecolumn (second format)
\smartqed  % flush right qed marks, e.g. at end of proof

\usepackage{amsmath}
\usepackage{amssymb}		% \mathbb{R} symbols
\usepackage{dsfont}			% \mathds{R} better than mathbb!
\usepackage{times}
\usepackage{graphicx}
\usepackage{placeins}
\usepackage{pdfpages}		% include pdf files
\usepackage{hyperref}		% clickable references
\usepackage[]{natbib}
\usepackage[english]{babel}
\usepackage[misc]{ifsym}	% letter symbol for corresponding author

\graphicspath{{./img/}{./plots/}}

\usepackage{adjustbox}
\hypersetup{
  colorlinks   = true,    % Colours links instead of ugly boxes
  urlcolor     = blue,    % Colour for external hyperlinks
  linkcolor    = blue,    % Colour of internal links
  citecolor    = blue     % Colour of citations
}

\usepackage{bbold}								
\begin{document}
\title{Estimating the extent of glioblastoma invasion}
\subtitle{Approximate stationalisation of anisotropic advection-diffusion-reaction equations in the context of glioblastoma invasion}
\author{Christian Engwer \and Michael Wenske}
\institute{
  C. Engwer \at
  Institut f\"ur Numerische und Angewandte Mathematik, WWU M\"unster, M\"unster, Germany \\
  \email{christian.engwer@uni-muenster.de}           %  \\
  \and
 M. Wenske \Letter (Corresponding author)\at
  Institut f\"ur Numerische und Angewandte Mathematik, WWU M\"unster, M\"unster, Germany \\
  \email{m\_wens01@uni-muenster.de}           %  \\   
}

\date{Received: date / Accepted: date}
% The correct dates will be entered by the editor

\maketitle
\begin{abstract}\noindent 
  Glioblastoma Multiforme is a malignant brain tumor with poor prognosis.
  There have been numerous attempts to model the invasion of tumorous 
  glioma cells via partial differential equations in the form of advection-diffusion-reaction equations.
  The patient-wise parametrisation of these models, and their
  validation via experimental data has been found to be
  difficult, as time sequence measurements are generally
  missing. Also the clinical interest lies in the actual (invisible)
  tumor extent for a particular MRI/DTI scan and not in a predictive
  estimate.
  Therefore we propose a stationalised approach to estimate the extent of 
  glioblastoma (GBM) invasion at the time of a given MRI/DTI scan. The
  underlying dynamics can be derived from an instationary GBM model,
  falling into the wide class of advection-diffusion-reaction
  equations.
  The stationalisation is introduced via an analytical solution of the 
  Fisher-KPP equation, the simplest model in the considered model class.
  We investigate the applicability in 1D and 2D, in the presence of 
  inhomogeneous diffusion coefficients and on a real 3D DTI-dataset.
   %Biomedical imaging and signal processing }
\end{abstract}
\section{Introduction}
\label{sec:introduction}
Treatment of glioblastoma multiforme (GBM) turmors usually consists of
a combination of tumor resection (operation), radio- and
chemotherapy \citep{SathoReard2007}. The treatment planning for this
type of tumor is particularly challenging, as the medical images do not show a clear
boundary and cancerous glia cells infiltrate seemingly healthy tissue far away from the visible
center, leading to a diffusive front. Tumor cells have been histologically cultivated from healthy
appearing tissue as far as 4 cm away from the bulk of the tumor \citep{SilberChic1997}. The
non invasive medical imaging modalities may only detect the tumor
upwards of a finite density threshold of about $16\%$
\citep{SwanRost2008,PateHath2017}, so that tissues are segmented as healty,
although they still contain a significant number of tumor cells.

In clinical practice an average extent of this invisible infiltration
of 2 cm normal to the visible tumor is assumed.
Using mathematical modelling the aim is to
estimate the extent of resection and radiotherapy to be applied
\textit{outside} of the tumorous regions visible on the medical images.
\paragraph{Modelling:}
Many efforts have been made to mathematically model the behaviour of
the tumorous glia cells within the brain. The mathematical approaches
are numerous and include a wide range of effects possibly influencing
the behaviour of the glioblastoma (GBM) cells. One prominent
mathematical approach is to describe the proliferation and the
movement of the tumor by macroscopic partial differential
equations. Most of these models take the form of
diffusion-reaction-advection equations.

Harpold et. al wrote a review article in 2007 about the evolution of
mathematical modelling of GBM \citep{HarpAlvo2007}. The modelling
started from simple reaction-diffusion equations with exponential
growth
($\frac{\partial c}{\partial t} = \nabla (\mathbf{D}\nabla c) + \rho
c$) \citep{TracCruy1995}.  From here, the extensions from
homogeneous diffusion to distinguished diffusivities in grey- and
white matter were performed by \citep{Swan2000}. With the advent of
diffusion tensor imaging (DTI) it was also possible to make use of
directional information of water diffusion within the
brain. \citet{Jbab2005} estimated the tumor diffusion from the water
diffusion matrices available from DTI and simulated the tumor invasion
making full use of the medical imaging information. It was also
possible to link the predicted spreading velocity of the Fisher model
($v\approx 2\sqrt{D\rho}$) to experimental data for low-grade
gliomas. They effectively found that the increase in radius was
linear. Thereby, giving further indication that the tumor growth may
be modelled by reaction-diffusion equations \citep{MandDela2003}.

This may indicate that the rate of advance can be estimated with the
knowledge of the diffusive properties of the surrounding tissue, and
the reproductive rate of the tumorous glia cells.  There are also
quite rigorous approaches to link knowledge on microscopic cell
behaviour to macroscopic partial differential equations via
upscaling. The mathematical framework was given by
\citep{OthmHillen2000,OthmHillen2002,HillPain2013}. The underlying
idea is to identify the most important influences on the cell's
behaviour on the microscopic and mesoscopic scale and mathematically
derive the influence on the macroscopic population dynamic. With this
powerful approach it has been possible to microscopically motivate
haptotaxis, chemotaxis and proliferation
\citep{Engwer2016a,Engwer2015a,HuntSuru2017,KelkSuru2011,Corb2018}.

Woodward et al. additionally modelled the effect of resection on the
tumor invasion, finding some correlation with real survival times of
patients \citep{WoodCook1996}.  A similar approach was taken by \citet{HuntSuru2017} on more advanced models. They modelled
the effect of chemotherapy on the ligand binding rate, the influence
of radiation killing cells and the resection itself. The resection was
simulated by depleting all tumor cell densities above a threshold
(e.g. 20\%). They compared different combinations of the simulated
treatment.  While this approach is more acknowledging to the
availability of data, the fundamental problems persist.

Currently, most models strive to describe the full temporal and
spatial dynamics of uninterrupted tumor growth but there also have
been been approaches to statically estimate the tumor's invasive
profile. Notably Konukoglu et al. formulated travelling-time
formulations for the tumor invasion problem in the form of eikonal
equations that only rely on the imaging data at the time of diagnosis
\citep{KonuClat2006,KonuSerm2007}. Their results are promising, but
the approach does not seem to not have found much traction.

\paragraph{Parameters:}
There have been approaches to assess the growth characteristics and
parameters in in-vitro experiments, e.g. \citep{OraiTzam2018}. It is
an open question whether the information gathered in these experiments
is transferable to in-vivo situations. Caragher et al. investigated
treatments using novel 3D cell culturing methods in the context of GBM
therapy developement \citep{CaraChal2019}.
% They state the following: \textit{"One key issue underlying this
% failure of therapies that work in pre-clinical models to generate
% meaningful improvement in human patients is the profound mismatch
% between drug discovery systems, cell cultures and mouse models - and
% the actual tumors they are supposed to imitate. Indeed, current
% strategies that evaluate the effects of novel treatments on GBM
% cells in vitro fail to account for a wide range of factors known to
% influence tumor growth."} \citep{CaraChal2019}.
Even if in-vitro experiments may prove essential in understanding the
involved biochemistry and qualitative effects, their use for the
validation problem is limited. In order to improve our ability to
accurately describe GBM invasion the interlinked problems of
parametrisation and availability of data have to be addressed. The
reason for this is that the large number of free parameters can
not be met with experimental data to estimate them with reasonable
accuracy.

\paragraph{Validation:}
\FloatBarrier
Although there are a number of mathematical descriptions available,
it has proven difficult to derive their parametrisation from
medical data or experiments.
In medical practice the diagnosis of GBM is often rapidly followed by
a combination of tumor resection, radio- and chemotherapy, thereby
severely altering the growth characteristics of the tumor
\citep{SathoReard2007}. One problem is, that in order to validate
any given forward GBM-model given in the form of
advection-diffusion-reaction equations, one would need a time-series
of DTI/MR/CT scans of the patient, where the tumor growth had been
left without cessation. That setup would allow for direct comparison
of the in-silico experiments and the medical images of the progressing
tumor. One also preferably had these datasets from a large number of
representative patients. In order to retrieve a dataset like that, one
needed to deprive a high number of patients of live prolonging
treatment while undergoing regular medical scans. The ethical
impossibility is obvious. It is also questionable whether any such
datasets will be available in the near future.
\begin{figure}
\begin{center}
\includegraphics[width=0.6\linewidth]{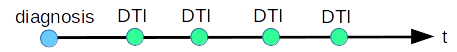}
\caption{Schematic time line of ideal datasets for the validation of 
GBM forward models.}
\label{fig:datasets_ideal}
\end{center}
\end{figure}
\begin{figure}
\begin{center}
\includegraphics[width=0.6\linewidth]{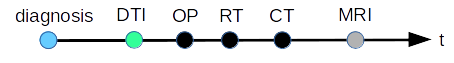}
\caption{Schematic time line of a realistic sequence of measurements. 
The initial DTI scan after diagnosis is rapidly followed by a 
combination of gross tumor resection (OP), radio- and chemotherapy (RT, CT).
 During treatment, there may be follow-up MRI's .}
\label{fig:datasets_real}
\end{center}
\end{figure}
Even in more accessible subjects like rodents, it proved difficult to determine a
good parametrisation for numerical models. Rutter et al. studied tumor
growth in five mice which were injected with tumorous glia cells under
controlled conditions. Even with a reportedly careful experimental
setup, the resulting tumor sizes varied significantly and fitting
parameters of a simple Fisher-KPP tumor model proved difficult
\citep{RuttStep2017}. Given the goal of improving the ability to
accurately describe the tumor invasion in real patients, the problem
of validation/falsification has to be addressed.

\paragraph{Contribution:}
We present a stationalisation approach, that opens the possibility to
estimate the invisible tumor extent at measurement time, building upon
existing instationary tumor growth models. It may also alleviate part of the parametrisation problem by
only being sensitive to the relative strength of physical effects and
not on their absolute quantitative parametrisation. We will first
state the class of considered partial differential equations in
section \ref{subsec:PDE}. In section \ref{subsec:stationalisation} we
present a methodology to find a stationalised analytical expression
for the 1D Fisher-KPP equation, determining the front shape of a
travelling wave solution. We also explain how the analytically derived
stationalisation term may be used to approximate the tumor
density. After numerical verification of the derived gradient
distribution, we investigate whether this stationalisation approach is
viable in the presence inhomogeneous material properties in section
\ref{sec:numerical_results}. Finally the advantages and shortcomings
of the proposed procedure will be critically discussed in
\ref{sec:discussion}.
\section{Modelling}\label{sec:modelling}
\subsection{Fully anisotropic advection-diffusion-reaction equation}\label{subsec:PDE}
The time-dependent GBM invasion is often modelled by parabolic partial differential equations. 
We describe the tumor density with the function $u : \Omega \times T \mapsto  \mathds{R}$,
with the  d-dimensional domain $\Omega \subset \mathds{R}^d$ and a
time range $T=[t_0, t_e]$. 
The solution $u$ describes the volume percentage of cancerous cells
and therefore $0\leq u(\mathbf{x}) \leq 1$.
The dynamic of the density profile is given in the form of a macroscopic partial differential equation. 
We consider the time-dependent fully anisotropic advection-diffusion-reaction equation in the following form:
\begin{subequations}
    \label{eq:myopic_diff_react}
  \begin{align}
\frac{\partial u}{\partial t} 	- \nabla(\mathbf{D}_t(\mathbf{x}) \nabla u)
                                - \nabla ((\nabla \cdot \mathbf{D}_t(x)) u )
  &=\rho u(1-u)
  &\text{in~}&\Omega\times T,\label{eq:myopic_diff_react_pde}
\intertext{with boundary and initial conditions}
  u(\mathbf{x},0) 				&= g(\mathbf{x})
  &\text{in~}&\Omega,\\
\nabla u(\mathbf{x},t) 			&= 0
  &\text{on~}&\partial\Omega\times T,
\end{align}
\end{subequations}
with the diffusion parameter $\mathbf{D}_t(\mathbf{x})$ being
a symmetric positive definite matrix $\mathbf{D}_t(\mathbf{x}) \in
\mathds{R}^{d \times d}$, i.e. the diffusivity can be inhomogeneous and anisotropic.
Equation \eqref{eq:myopic_diff_react} is a prototype model for the invasion of GBM in the sense that many
models differ from it merely in the reconstruction of the tumor diffusion matrix $\mathbf{D}_t(x)$ from DTI data,
or the addition of other chemotaxis terms \citep{HuntDiss2018}.

\subsection{Stationalisation of the Fisher-KPP equation}\label{subsec:stationalisation}
In one dimension and for isotropic diffusive properties with ($\mathbf{D}_t = 1$, $\rho = 1$),
equation \eqref{eq:myopic_diff_react} degenerates to the classical
Fisher-KPP equation \citep{fisherKPP1988,Fisher1937};
\begin{equation}
\frac{\partial u(x') }{\partial t} = \Delta u(x') + u(x')(1-u(x')),
\label{eq:fisher}
\end{equation}
with $x' \in \mathds{R}$ as a spatial coordinate. The first term may be interpreted as the passive diffusive spread of the cells due to a random walk, 
the rightmost term as a logistic proliferation term, as often encountered in biological contexts. 
For the initial conditions:
\begin{equation*}
\lim\limits_{x' \rightarrow -\infty}{u'(x) = 1} ,~~\lim\limits_{x' \rightarrow \infty}{u'(x) = 0} , 
\end{equation*}
the equation allows travelling wave solutions \citep{fisherKPP1988}. 
These solutions converge over asymptotically long times cales 
to a wave-front which is constant in shape, moving laterally with a globally constant velocity $v \in \mathds{R}$.
With that information, equation \eqref{eq:fisher} can be stated in 
equivalent form by introducing a co-moving frame $x = x'-vt$.
\begin{equation}
0= \Delta u(x) + u(x)(1-u(x)) \underbrace{- v \cdot \nabla u(x)}_{\text{ stationalisation}}
\label{eq:fisher_comoving}
\end{equation}
The advective term $- v \cdot \nabla u(x)$ results from the coordinate transformation into 
the comoving frame and can be understood as simply counteracting the lateral movement of the propagating front \citep{Ablo1979,fisherKPP1988}. 
We will refer to the advective term and its approximations as the
\textit{stationalisation term}.
Any laterally shifted profile $u(x-c)$, with a constant $c \in \mathds{R}$ is equally 
a solution to \eqref{eq:fisher_comoving}.
The minimum limit speed of the travelling wave solutions has been stated to be
\begin{equation*}
v = 2\sqrt{D_t \rho~}\!,
\end{equation*}
but Kolomogorov also states that there are solutions for speeds higher than $v$.
This fact is due to the reaction term acting independently of the local environment.
For the limit of diminishing diffusion, the reaction term will induce very high travelling wave speeds
for initial conditions which are close to horizontal. Contrarily, the wave speed will be determined
to a higher degree by the diffusive flux for initial conditions that fall from 1 to 0
in a very localized region \citep{fisherKPP1988}. In the context of glioma invasion
only the very rapid spatial decay of tumor density is relevant.
Even though the sigmoidal shape of the advancing front profile is formed quite rapidly, 
it may take asymtotically long to reach a state where 
the shape of the advancing front is no longer changing.
For long time scales and no boundary conditions, 
a stationary wave form can be given in analytical form for the 
special wave-speed of $v = \frac{5}{\sqrt{6}}$
\begin{equation}
u(x) = \frac{1}{(1- r \exp(\frac{x}{\sqrt{6}}))^2}~,
\label{eq:fisher_sol}
\end{equation}
where $r =-1$ \citep{Ablo1979}.
We will call that profile the \textit{equilibrated} wave-form. 
In the fully equlibrated case, we can state an analytical expression for $\nabla u(x)$. 
Since the analytical solution is invertible in the relevant range $u \in [0,1]$, 
we may also express the gradient in terms of $u$.
Assuming an analytical expression for $p(u)= v\nabla u (u)$, we may 
express the comoving 1D Fisher-KKP equation
\eqref{eq:fisher_comoving} in equivalent form:
\begin{equation*}
0= \Delta u + u(1-u)\underbrace{- p(u)}_{\text{ stationalisation}}.
\end{equation*}
The gradient of the analytical solution, can be given as
\begin{equation}
\nabla u= \frac{ -\sqrt{\frac{2}{3}} \exp(\frac{x}{\sqrt{6}})    } 
{(1 + \exp( \frac{x}{\sqrt{6}} ))^3}.
\label{eq:sol_grad}
\end{equation}
The inverse of the analytical solution is given by
\begin{equation}
x(u)= \sqrt{6} \ln \Big(\pm\frac{1}{\sqrt{u}} - 1\Big).
\label{eq:fisher_sol_inverse}
\end{equation}
and represents the mapping from a given amplitude $u$ 
to the corresponding position $x$ within the wave form profile. 
Substituting the inverse \eqref{eq:fisher_sol_inverse} into the gradient expression\eqref{eq:sol_grad} yields
an analytical term for $\nabla u(u)$ and thereby a closed form
for the stationalisation term
\begin{equation*}
p(u)  =|v| \sqrt{\frac{2}{3}} (1-\sqrt{u})u.
\label{eq:fisher_sol_dudt_of_u}
\end{equation*}
\emph{Note} that the strict equivalence between $p(u)$ and $-v \nabla u$
 only holds for the limit case, in which the analytical solution is available. 
Also, the expression only holds for the case without
boundary conditions, i.e. $\Omega \equiv \mathds{R}$.
Although this formulation is only strictly equivalent in case of a completely 
equilibrated wave form, we will use the penalty term $p(u)$ as an \textit{approximation} of $v \nabla
u(u)$ leading to a stationalised wave-pinning type model:
\begin{subequations}
  \begin{align}
    \label{eq:fisher_pen}
    0 &= {\underbrace{\Delta u  + 	u (1-u) \vphantom{
        \sqrt{\frac{2}{3}} }}_{\text{tumor model prototype}}}- {
        \underbrace{ |v|~ \sqrt{\frac{2}{3}}
        (1-\sqrt{u})u}_{\text{stationalisation}}} 	& \text{ in } \Omega,\\
         \nabla u(\mathbf{x}) &=0  					& \text{ at } \partial\Omega, \\
    u(\mathbf{x}) &=c  								& \text{ at } \partial\Omega_i .
  \end{align}
\end{subequations}
Within the encompassing domain $\Omega$ we define a smaller 
enclosed domain $\Omega_{i} \subset \Omega$ representing the tumor visible on the medical images.
The additional internal boundary condition on $\partial \Omega_i$ is used to localize the stationalised solutions.

The Fisher equation is one example out of a family of KPP-type equations 
which combine a diffusive term with a nonlinear reaction term $f(u)$.  
The reaction term is often chosen in a manner so that it dynamically 
connects two fixed points of amplitude: $f(0)=0, f(1)= 0, f(0<u<1)>0$. 
Although an exact analytical description of the stable wave
 fronts proves difficult, the rough characteristic of propagating fronts 
 found in nature (combustion, bacteria growth etc.) is often similar to a sigmoid function. 
The gradient distribution of any sigmoid-like travelling wave front will have a 
similar shape like $p(u)$. For any sigmoidal wave-form $|\nabla u|$ will
 be zero for $u=0$ and $u=1$ and of higher amplitude for
$0<u<1$.

The underlying idea of the stationalisation procedure is, 
that the gradient distribution, and therefore the penalty term necessary to
calculate the travelling wave form stationally, may possibly be approximated to a good 
degree of accuracy and even for those cases where the analytical solution is not at hand. 
Considering the general model in the form of
\eqref{eq:myopic_diff_react}, we define the corresponding stationalised
problem as: Find $u_s$ such that
\begin{subequations}
  \begin{align}
0&=\nabla(\mathbf{D}_t(\mathbf{x})\nabla u_s)+ \nabla((\nabla \cdot \mathbf{D}_t(\mathbf{x})) u_s ) \notag\\
&\phantom{=} + \rho u_s(1-u_s)- |v|~ \sqrt{\frac{2}{3}} (1-\sqrt{u_s})u_s,&\text{ in }\Omega,    \label{eq:stationary_fisher_with_D}\\
         \nabla u(\mathbf{x}) &=0  			&\text{ at } \partial\Omega,~ \\
    u(\mathbf{x}) &=c  						&\text{ at } \partial\Omega_i .
  \end{align}
\end{subequations}
The above problem closely resembles the general model \eqref{eq:myopic_diff_react_pde} augmented with the
penalty term $p(u)$ and internal boundary conditions.
The matrix $\mathbf{D_t}(\mathbf{x})$ is a reconstruction of the tumor diffusivity from the DTI datasets, 
$\rho$ is a growth parameter and $v$ the penalty parameter.

It is known that in higher spatial dimensions, the Fisher-KPP equation has related sigmoid-like travelling wave solutions. 
There are more additional stable wave front patterns in higher dimensions. 
One of them describes a v-shaped waveform propagating through the two-dimensional medium, 
which can be interpreted as two straight wave fronts collapsing into each other at a certain angle. 
Observed in the direction of a bisecting line, this combined wave front is indeed stationary at certain speeds. 
There also exist spatially oscillating front shapes, but these profiles are 
only possible for the extension of $u(x)$ out of the relevant range $u \in [0,1]$ \citep{BrazTyson1999}. 
We focus on cases where the wave propagation occurs as a radial expansion from a centred mass. 
In the direction of the propagation, equation 
\eqref{eq:fisher_sol} provides good estimates on the wave fronts profile.

\section{Numerical methods and error measures}
\label{sec:numerical_methods}

In this section we discuss the methods used in our numerical
experiments to measure the modelling error of the proposed
stationalisation.
In section \ref{sec:numerical_results} we will use the methods to
investigate the validity of our approach in a series of problems with growing complexity.

\subsection{Numerical scheme}\label{subsec:numerical_method}
We follow the method of lines approach to split temporal and spatial operators.
The temporal discretisation is an implicit-euler scheme, which is
unconditinally stable.

For the spatial discretisation we use for simplicity a standard
finite element discretisation on cubic grids with multi-linear trial-
and test-functions.

The matrix divergences are pre-calculated by a first order finite
difference approximation within each grid cell. The inhomogeneous
diffusion matrices at the quadrature point are evaluated by nearest
neighbour interpolation.

Positivity of our solution might be violated in finite precision
calculations. Wherever the numerical iteration leaves the
physically sensible range of $u \in [0,1]$, we disregard the
reaction term of the given forward- or stationalized model and
instead employ the following artificial numerical penalty term
\begin{equation}
n(u)=
\begin{cases}
-\rho u 	&\text{if~} u<0\\
\rho (1-u) 	&\text{if~} u > 1.\\
\end{cases}
\label{eq:reaction_with_oob_pen}
\end{equation}
This is done, because a logistic reaction term $ \propto \rho u(1-u)$ may otherwise amplify
numerical fluctuations which produce a slightly negative
amplitude in $u$.

The stationalisation procedure should be largely independent of the
chosen numerical discretisation. The largest source of error does not
lie in the numerical treatment, but in the approximations made in the
parametrisation and in the estimation of the internal Dirichlet
constraints.

\subsection{Implementation}
The implementation was realized within DUNE software framework \citep{dunegrid1,dunegrid2,dune24:16}. 
The finite element discretisation was implemented within dune-pdelab
\citep{dune-pdelab}. The non-linear system is solved with a classical
Newton-Krylov method, using linear search. The linear systems
are solved with an AMG- preconditioned BiCGSTAB solver, using the dune-istl module \citep{BICGSTAB,dune-istl}.
The release version for all DUNE modules was 2.6.

For the realistic head model the diffusion matrices were reconstructed by the camino software package \citep{camino}.

\subsection{Error measure}
\label{subsec:errormeasure}
In section \ref{subsec:effect_on_front} we will investigate the impact of the
stationalisation error on the observed tumor front.
In this course we will compare the reconstructed tumor front of a
fully instationary simulation with the reconstructed tumor front using
the stationalisation approach.

\begin{figure}[htp]
\centering
\includegraphics[width=0.25\linewidth]{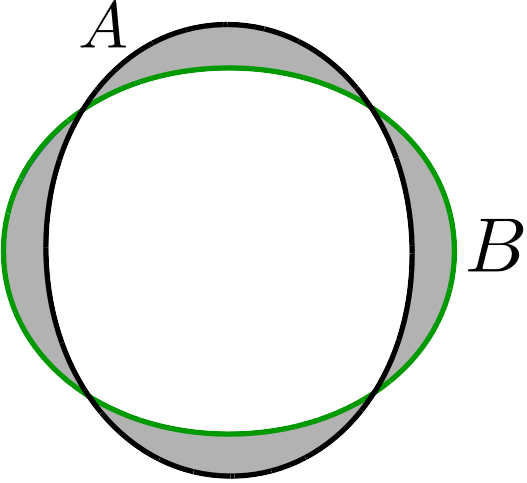}
\caption{Schematic representation of $A \oplus B$: grey regions}
\label{fig:error_venn}
\end{figure}

Given a reference solution $u_a$, an approximation $u_b$ and a threshold
value $\theta$ we define two domains $A$ and $B$ as
\begin{equation*}
  A = \{\mathbf{x} | u_a(\mathbf{x})\geq \theta\},\qquad 
  B = \{\mathbf{x} | u_b(\mathbf{x})\geq \theta\}.
\end{equation*}
The medically relevant information is the spatial discrepancy between
two level-sets ($\partial A, \partial B$) of these density
profiles. An absolute measure for this error is the symmetric
difference $A \oplus B$, as depicted in Figure \ref{fig:error_venn}. It
measures those sub-volumes 
which are either included $A$ but not in $B$, or vice versa. 
That way, both over- and underestimations of the approximation $u_b$ are represented.
The most expressive information in the medical context might be the
average distance between the two level-sets.
We therefore introduce the characteristic level-set distance.
\begin{definition}[Characteristic level-set distance]
  For a given level-set value $\theta$,
  we define the characteristic level-set distance between $\partial A$
  and $\partial B$
  as
  \begin{equation}
    L_B :=  \frac{|A \oplus B|}{|\partial A|}.
    \label{eq:L_B}
  \end{equation}
  It quantifies the average distance between the 
  
  Assuming a spherical reference geometry for $A$ we can simplify this
  expression and avoid evaluating $|\partial A|$. Given the radius
  $r_A$ (or an approximate average radius of $A$), $L_B$ simplifies to
  \begin{equation*}
    L_B^\text{1d} = \frac{|A\oplus B|}{2}, \quad
    L_B^\text{2d} = \frac{|A\oplus B|}{2 \pi r_A} \quad \text{and}\quad
    L_B^\text{3d} = \frac{|A\oplus B|}{4 \pi r_A^2}.
  \end{equation*}
\end{definition}

\subsection{Notes on Uniqueness}
\begin{figure}[h!]
  \includegraphics[width=0.5\linewidth]{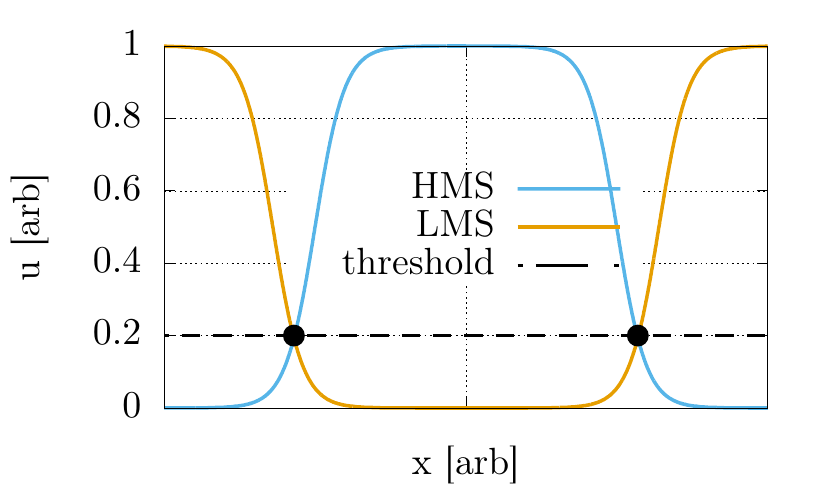}
  \includegraphics[width=0.5\linewidth]{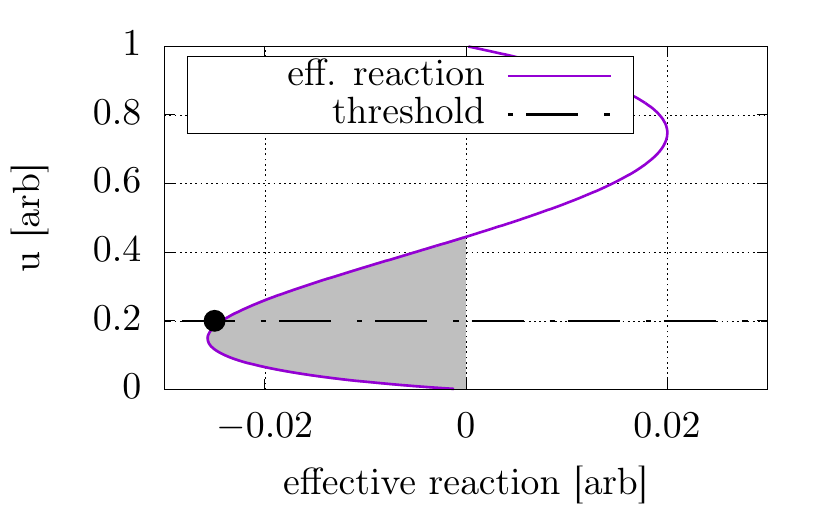}
  \caption{\textbf{Left:} Schematic of high- and low mass solutions (HMS, LMS), both fulfilling the
internal Dirichlet constraints (black dots). \textbf{Right:} Plot of effective 
  reaction term consisting of the logistic growth and the penalty term. 
  The penalizing regime is indicated in grey. The visibility threshold is within the penalizing regime.}
\label{fig:solutions_with_eff_react}
\end{figure}

\FloatBarrier
The limit solutions to the Fisher equation \eqref{eq:fisher} allow for travelling wave solutions
moving in both the positive and negative direction. In the stationalised (co-moving) formulation
we find a similar situation. If equation \eqref{eq:fisher_comoving} is augmented with a Dirichlet
side condition in the form of $u(\mathbf{x_c})=c$ within the domain and Neumann boundaries at the
border, we may find two solutions mirrored around the
constrained point within the profile. Since we want to use an internally constrained region
given by segmentation information we may expect two possible solutions
to the formulation \eqref{eq:fisher_comoving}.
The first solution corresponds to the stationary approximation to the travelling wave 
solution of the forward model, which moves outwards
from the tumor center, while satisfying the Dirichlet constraint. This solution will have
relatively low mass outside the constrained region. The second possible solution corresponds
to a travelling wave moving into the constrained region, and will have a high mass on the outside.
The two solution types are sketched in Figure \ref{fig:solutions_with_eff_react}.
When considering the effective reaction term consisting of the logistic growth combined with the penalty term,
we find that it has a penalizing regime for $0<u<\frac{4}{9}$ and
 a growth regime for $\frac{4}{9}<u< 1$ (Fig. \ref{fig:solutions_with_eff_react}).
The diffusive process transports mass from high amplitudes to lower amplitudes. The combined reaction term
 counteracts this process.
The visibility threshold, and therefore the constraints, lie within the penalizing regime.
In order to select the low mass solution branch, we use initial guesses $u_s(x) \ll 1$ which are well within
the purely penalizing regime, so that the newton iteration converges to the low mass solution reliably.

\section{Numerical results}\label{sec:numerical_results}
We present the results of the numerical validation studies of the
stationalisation procedure in 1D and 2D.
We first compare the gradient distributions of forward simulations with the analytical 
expression found in equation \eqref{eq:fisher_sol_dudt_of_u} in section \ref{sec:modelling}. 
After that we compare forward simulations
with their stationary approximations and assess the impact of our
modelling error on the reconstructed tumor front. We present an easily reproducible 2D example in \ref{subsubsec:2D_inhom_stat}.
Finally we present the applicability to a realistic 3D DTI dataset in \ref{subsec:3D_real_data}.

\subsection{Investigation of modelling error}
\label{subsec:validate_gradient_dist}
We will numerically investigate the effect of inhomogeneous diffusion
on the gradient distribution, and by this the applicability of
the chosen stationalisation term. As the analytic formulation of
$\nabla u$ is only valid in the homgeneous 1D case, we expect a
modelling error, which we will assess in different test cases.

\subsubsection{Artificial imhomogeneous diffusion}\label{subsubsec:inhom_diff_def}
We introduce a test cases with randomly 
perturbed diffusion coefficients.
\begin{equation}
\mathbf{D}_\beta = \mathds{1}_d  ~ (\delta)^{\frac{1}{d}},\label{eq:D_delta}
\end{equation}	
with $\delta$ being a uniformly distributed random value with a spread of $\beta$ around an average of 1.
Here, $d$ is the spatial dimension.
The exponent of $\delta$ is chosen to allow comparisons between the isotropic homogeneous case and the randomized case, 
by assuring that the average of the determinants of the 
diffusive medium are close to 1.0 for every realisation of the random field:
\begin{equation}
|\bar{\mathbf{D}}_\beta| = \bar{ (\delta)} \approx 1.
\label{eq:det_of_random_D}
\end{equation}
We only compare the homogeneous and inhomogeneous case with equal grid resolution.
We evaluate the random inhomogeneous diffusion on the dual grid and assume it to
be piecewise constant therein. The diffusivity within one dual grid cell is
statistically independent from any neighbour.
The effect of statistical scattering of the diffusive properties on the macroscopic front speed is non-trivial. 
Since the global front speed appears as a linear factor in the analytic derivation of the stationalisation term, 
we may not expect perfect convergence to the analytically derived gradient distribution. 
The results may however illustrate the effect of realistic datasets.

\subsubsection{1D gradient distributions}\label{subsubsec:1D_gradient_dist}
We consider a one dimensional forward simulation of equation \eqref{eq:myopic_diff_react}
starting from a Gaussian initial condition in the center of a one dimensional domain $x \in [0,200]$. 
We first simulate the homogeneous case with $D_t = 1$ where we expect perfect 
convergence of the gradient distribution to the analytical expression. 
In the homogeneous case there are no advective terms active, as $\nabla \cdot D_t$ is zero. 
Upon start of the simulation, two travelling waves form and move away from the center of the domain. 

\begin{figure}
  \centering
  \includegraphics[width=0.8\linewidth]{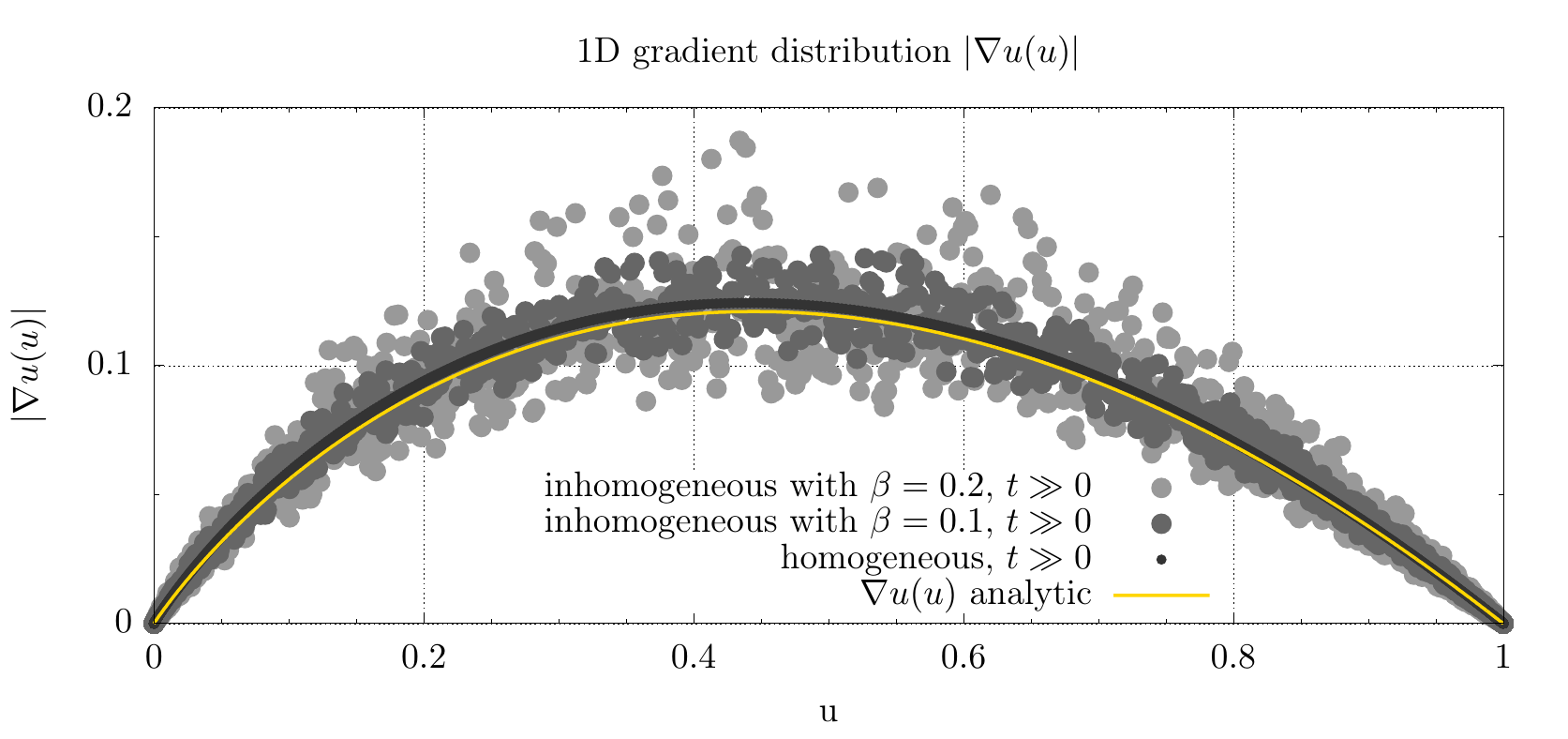}
\caption{Scatter plot of 1D gradient distribution: magnitude of gradient of 
1D wave-front profile. The numerical solution approaches the analytic 
expression only at asymptotically large time-scales. The underlying 
characteristic is not destroyed by the introduced inhomogeneities.}
\label{fig:1D_inhom_grad}
\end{figure}

After a short initial phase, the wave-fronts in the homogeneous medium asymptotically 
approach the front shape of the analytical solution \eqref{eq:fisher_sol} and its symmetric counterpart. 
The front speeds within an inhomogeneous material are slightly
perturbed. Similarly the relation $\nabla u (u)$ is disturbed.

In Figure \ref{fig:1D_inhom_grad} we investigate how the gradient distributions approach the analytical 
expression for the homogeneous and inhomogeneous case and how strongly
it differs from the analytical expression in the case of heterogeous coefficients.

\subsubsection{2D gradient distributions}\label{subsubsec:2D_gradient_dist}
We consider a square domain ($L_x = L_y=200$) in 2D with 
a Gaussian initial condition in the center. 
After the start of the simulation, the wave-front circularly propagates 
outward from the central point. 
Slow convergence of the gradient distribution to the analytic 
expression \eqref{eq:fisher_sol_dudt_of_u} can be replicated in the 2D homogeneous case. 
\begin{figure}[hbt]
  \centering
  \includegraphics[width=0.8\linewidth]{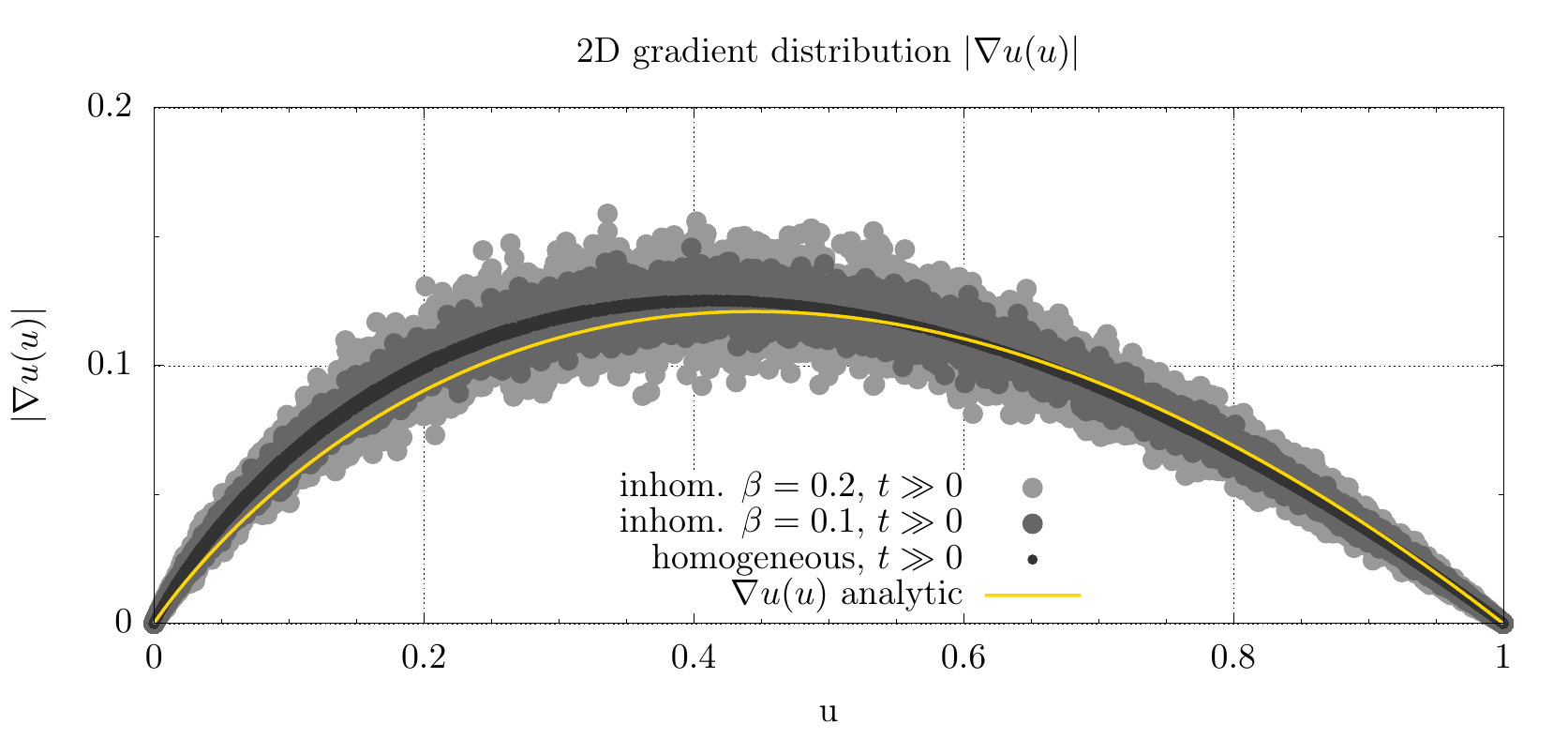}
\caption{Gradient distribution of a 2D simulation with different 
spreads in diffusive properties ($\beta = 0.0, 0.1, 0.8$).}
\label{fig:2D_inhom_grad}
\end{figure}
Contrary to the one dimensional case, there is an effect of curvature in higher dimensions
slowing the convergence towards the 1D gradient distribution.
However in the limit case the wave propagation still reduces to a 1D dynamic
in the propagation direction \citep{Murray2007}.
It is obvious that the introduction of random material properties breaks 
the strict applicability of the stationalisation term. 
However Figure \ref{fig:2D_inhom_grad} 
suggest that also in 2D the underlying polynomial relation between the wave-fronts 
amplitude and its gradient is merely perturbed by the material properties. 
Compared to the usual parameter uncertainties, we consider this
modelling error small and thus a stationalised solution, making use of the 
analytical gradient distribution, should still provide reasonable estimates 
on the density profile. Since the global propagation speed $v$ appears as a 
linear factor, any process that alters the propagation speed away from the 
analytical value should have a linear effect on the gradient distribution.
% \FloatBarrier

\subsection{Effect on estimation of the tumor front}\label{subsec:effect_on_front}
The stationalisation includes a modelling error due to
the imperfect approximation of $\nabla u$. The numerical observation in section
\ref{subsec:validate_gradient_dist} suggests that the average behavior
is still well described by our analytic reformulation \ref{eq:fisher_pen}. 
In the following we will investigate the impact of
this stationalisation error on the actual tumor front. We will use
test cases with growing complexity.

\begin{figure}[htbp]
  \centering
  \includegraphics[width=0.45\linewidth]{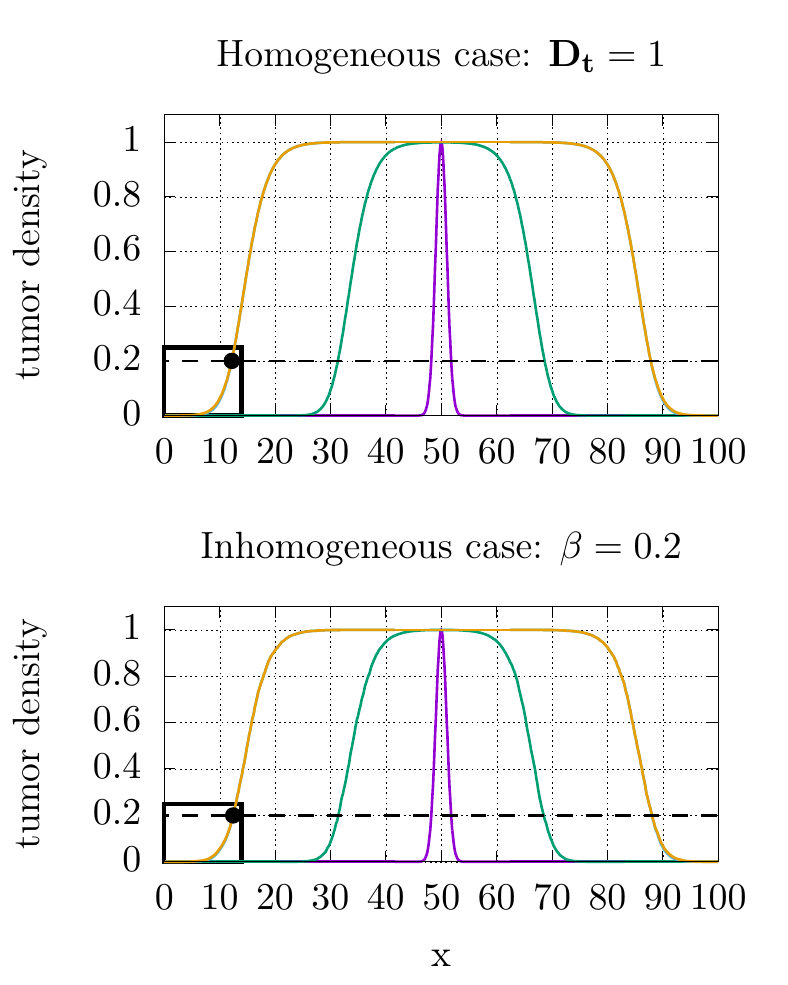}
  \includegraphics[width=0.45\linewidth]{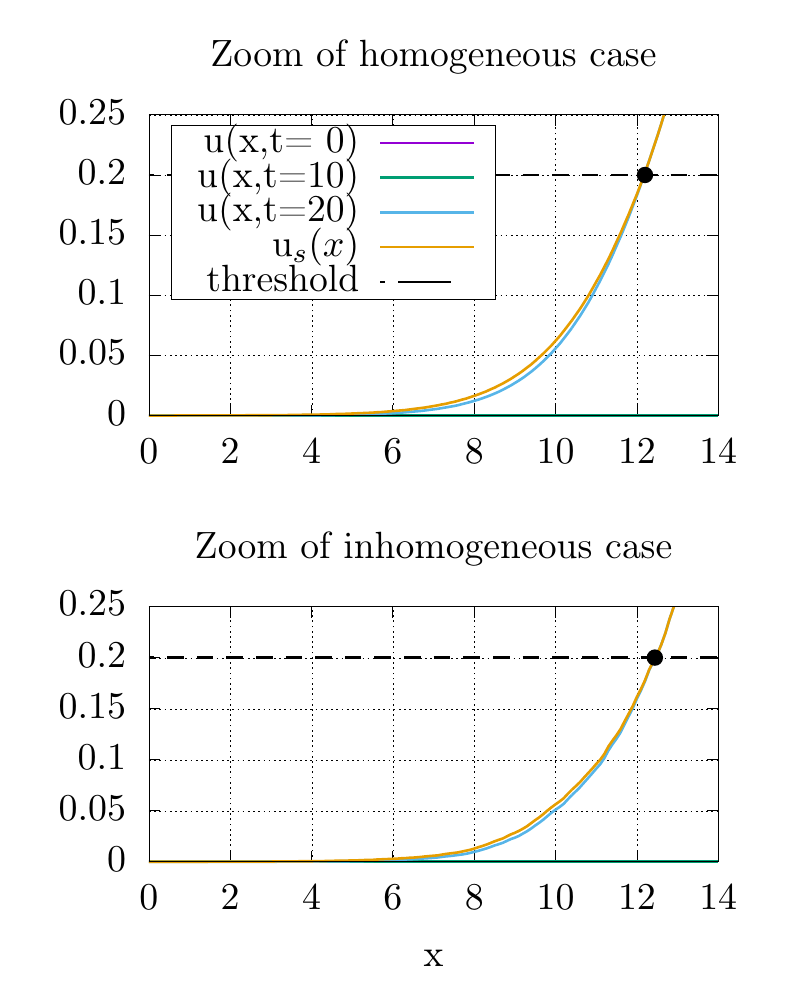}
  \caption{Direct comparison of 1D forward simulations and their stationalizations. The black dots indicate the begin of the internal constraint given by the simulated imaging threshold ($u=0.2$).
 The domain was discretised into 1000 equidistant elements. \textbf{Top left.:} Initial condition, and states of the forward simulation (homogeneous).
  \textbf{Bottom left.:} Initial condition, and states of the forward simulation (inhomogeneous).
  \textbf{Top right.:} Zoom of the forward solution and the corresponding stationalization (homogeneous).
  \textbf{Bottom right.:} Zoom of the forward solution and the corresponding stationalization (inhomogeneous).
   Both the forward and the stationalised solution 
show slight deviations from a smooth decay in the inhomogeneous case. Note that the front speeds are not
   perfectly identical.
}\label{fig:1D_combined}
\end{figure}

We try to mimic the situation observed in the medical application and present
the procedure of estimating the tumor extent at the time of diagnosis.
To generate artificial datasets in controlled scenarios, we simulate the
 carciogenesis by first assuming a small Gaussian initial condition. 
Secondly, we propagate the density profile for a certain time 
simulating the uninterrupted tumor growth. 
Finally, at the time of diagnosis, we use a level-set 
of $u = 0.2$ to represent the thresholded medical imaging modalities. 
Other choices of threshold value are possible.
We then use the thresholded volume as an internal Dirichlet constraint and solve 
the stationary problem \eqref{eq:stationary_fisher_with_D}.
In this numerical setup both the full forward density 
profile $u$ and the stationary profile $u_s$ are known and compared. 
In any real world situation only the thresholded image information would be accessible. 

\subsubsection{1D front reconstruction}\label{subsubsec:1D_stat}
We first show results for a simple 1D case with $\mathbf{D}_t =1$ 
and with inhomogeneous coefficients ($\beta = 0.2$).
By introducing inhomogeneous diffusive coefficients, the 
advective drift term will produce small contributions to the equation.
We chose the penalty parameter $v= \frac{5}{\sqrt{6}}$ and $\rho = 1$.
The one dimensional setup is not very realistic, 
but practical to illustrate the procedure.
% \FloatBarrier
Figure \ref{fig:1D_combined} presents a direct comparison of forward
 simulations with their stationalized counterparts. The snapshots at different times of the
forward solution suggest, that the rough form of the advancing front is formed rapidly.
If the forward solution converges rapidly towards the form of the analytical solution, with
only diminishingly small corrections to the wave fronts shape at larger times, then correspondingly the
approximation to the gradient statistic will perform well even early in the simulation.
The direct comparisons show that the stationalization produces 
a reasonable approximation to the full forward solutions.
The inhomogeneous coefficients induce small deviations from
a smooth front shape, which are present in both the forward
simulations profile as well as in the stationalized solution.
It is to be expected that wherever the internal constraint results from 
thresholding of an underlying smooth distribution, 
the stationalisation will perform well since the real density 
distribution is close to the equilibrated wave-form. 

\subsubsection{2D butterfly test case}\label{subsubsec:2D_inhom_stat}
In order to assess the viability of the stationalisation for more realistic 
tumor models we now move to two dimensions with inhomogeneous coefficients 
(eq. \eqref{eq:stationary_fisher_with_D}). We set up an inhomogeneous but 
isotropic field for the diffusion matrix by scaling the unit 
matrix according to its $x_1$ position
\begin{equation}
\mathbf{D}_t(\mathbf{x}) = \mathds{1}_2~(1.0+\sin\Big(\frac{3 \pi}{L_x} x_1\Big)~0.9).
\label{eq:butterfly_D}
\end{equation}
This effectively separates the domain in three distinct regions, 
with the left- and rightmost third of the domain having higher diffusivity 
and the middle strip having reduced diffusivity. The changes in diffusivity 
may represent grey and white matter regions in a primitive way. In this
example there is more long-range deviation of the diffusive properties
than in the 1D examples in section \ref{subsubsec:1D_stat}.
We again chose the penalty parameter $v= \frac{5}{\sqrt{6}}$ and $\rho = 1$. 
In a medical situation the task is to estimate the region and intensity 
of radiotherapy to be applied and the area of resection 
from only the thresholded information visible at the time of diagnosis.

\begin{figure}[htbp]
  \centering
  \includegraphics[width=\linewidth]{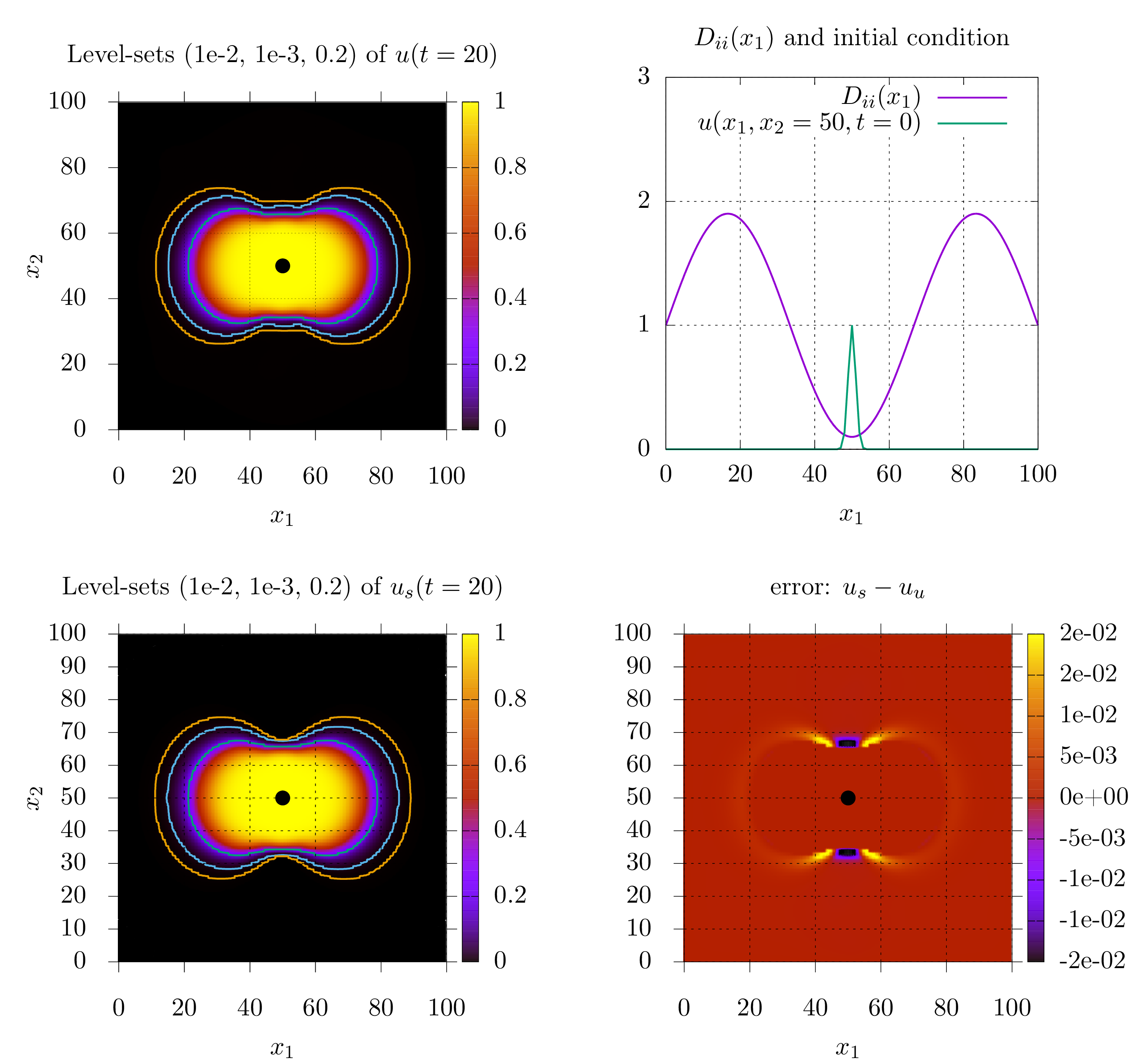}	%smaller total filesize for submission
\caption{2D inhomogeneous isotropic testcase.
 \textbf{Top left:} level-sets on the forward solution at t=20. 
 \textbf{Top right:} diffusivity as given in \eqref{eq:butterfly_D} and 
 horizontal cut through the Gaussian initial condition at $\mathbf{x}=(50,50)$.
 \textbf{Bottom left:} level-sets on the solution of the stationary problem. 
 \textbf{Bottom right:} Error field indicating 
 regions of over- and underestimation.}\label{fig:butterfly}
\end{figure}

Figure \ref{fig:butterfly} compares predictions of the forward simulation 
of equation \eqref{eq:myopic_diff_react} with the stationary solutions of 
equation \eqref{eq:stationary_fisher_with_D}. The stationalisation greatly
 benefits from the internally constrained region. Since most of the tumor 
 mass is above the visibility threshold, the stationalization only has to
 provide the estimation on the surrounding region.
Similar to the 1D constrained situation, the stationalisation captures 
the profile quite well, but slightly overestimates the invasion extent 
in low density regions.

%\FloatBarrier
\subsection{3D stationalisation for a realistic dataset}\label{subsec:3D_real_data}
To show the applicability on real patient data, we use the publicly accessible 
DTI-dataset provided by the camino\footnote{\url{http://camino.cs.ucl.ac.uk/index.php?n=Tutorials.DTI}}
software project \citep{camino}. We relate the tumor diffusion to the water 
diffusion by a simple scalar factor:
\begin{equation}
\mathbf{D}_t = \alpha \mathbf{D}_w.
\label{eq:D_t}
\end{equation}
More advanced reconstructions are possible, 
but not central to this example. A variety of different tumor
diffusion models has been proposed in the literature, see e.g. \citep{HuntDiss2018,PainHill2013,ConteGera2020}.
\begin{table}[t]
\begin{center}
\begin{tabular}{l|c}
\textbf{parameter} & \textbf{value }\\
\hline
$\alpha$ & 5e-12 \\
$\rho$ & 1e-6 [1/s]\\
$v$ & 2.04e-6 [1/s]\\
\end{tabular}
\caption{Parameters used to scale the terms in \eqref{eq:myopic_diff_react}
and \eqref{eq:stationary_fisher_with_D} to realistic ranges. The 
penalty parameter $v$ has the unit of $1/s$ since we approximated the 
gradient within the advective term with the 
analytical expression \eqref{eq:fisher_sol_dudt_of_u}.}
\label{tab:parameters}
\end{center}
\end{table}
We use the forward model \eqref{eq:myopic_diff_react}
and the corresponding stationary problem \eqref{eq:stationary_fisher_with_D},
with the parameters in table \ref{tab:parameters} to scale the equations to a realistic range.
Here $\alpha$ is a dimensionless parameter, $v$ the penalty parameter,
and $\rho$ is a growthrate in $1/s$.
We again follow the procedure described in section \ref{subsec:effect_on_front} 
and start a forward simulation from a small Gaussian at $t_0 = 0$ until 
$t_e = 90d$, and use a level-set ($u=0.2$) as the constrained region for the stationalisation.
\begin{figure}
  \centering
  \includegraphics[width=\linewidth]{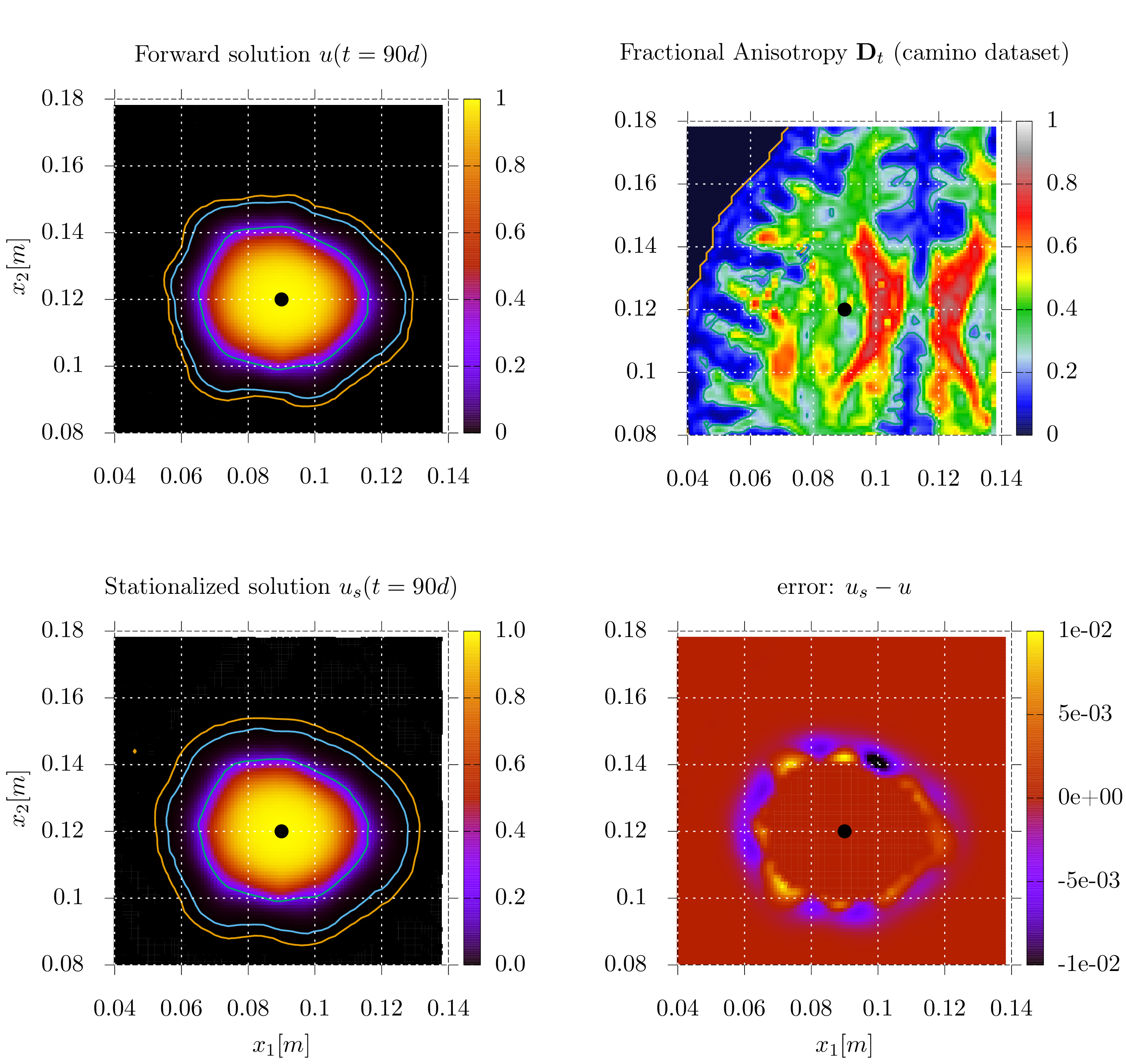}	%smaller filesize for submission
  \caption{Horizontal slice of the 3D results and dataset. 
    The black dot indicates the position of the small Gaussian initial 
    condition at $x=	(0.09m,0.12m,0.05m)$. 
    \textbf{Top left:} level-sets 
    (0.2, 1e-3, 1e-4) on $u(\mathbf{x})$ after 90 days. 
    \textbf{Top right:} Fractional Anisotropy of the reconstructed tumor 
    diffusion matrix $\mathbf{D}_t(x)$ from the 
    camino dataset \citep{camino}.
    \textbf{Bottom left:} Identical level-sets on $u_s(\mathbf{x})$.
    \textbf{Bottom right:} Regions of over- and underestimation by the stationary solution.
  }
  \label{fig:3D_real_data}
\end{figure}
\begin{figure}
\begin{center}
    \begin{adjustbox}{width=\linewidth}
      \includegraphics{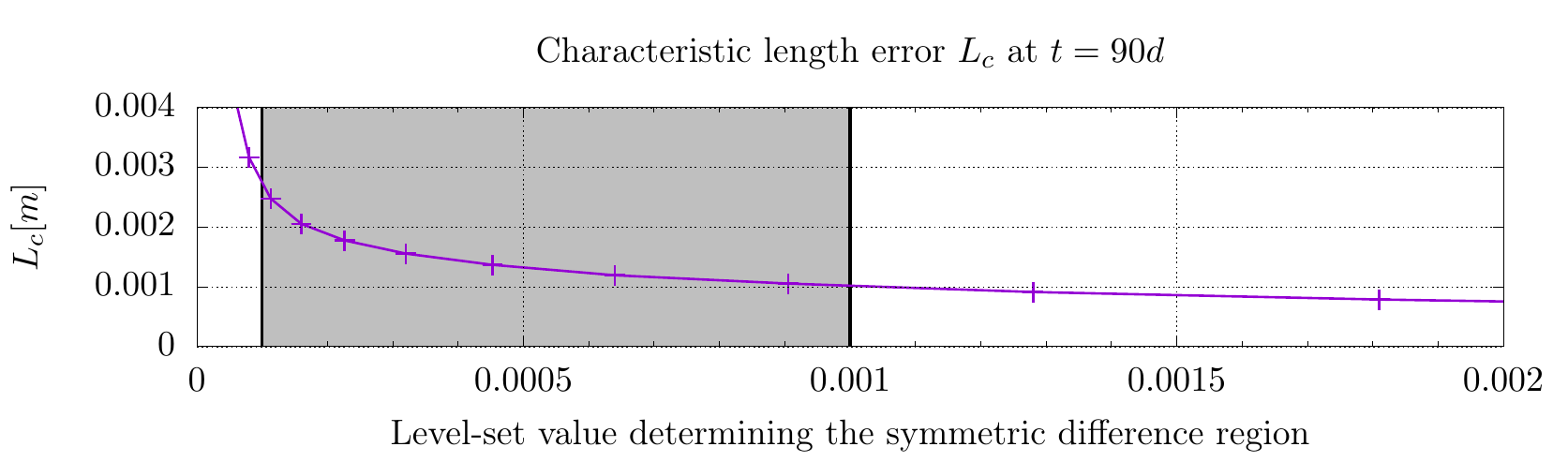}
    \end{adjustbox}
\caption{Characteristic length errors $L_B$ for given level-set values 
for the 3D example \href{http://camino.cs.ucl.ac.uk/index.php?n=Tutorials.DTI}{camino} 
dataset. The grey region indicates the errors between the level-sets used in Figure \ref{fig:3D_real_data}.}\label{fig:length_error}
\end{center}
\end{figure}
Figure \ref{fig:3D_real_data} shows the direct comparison of the forward simulation and the stationalization.
All local extentions or reductions induced by the local increase or decrease of the underlying
diffusivity are present in both the forward and the stationalized solution.
For this particular example, we measured the characteristic levelset distance $L_B$
as given in \eqref{eq:L_B} for a series of small levelset values.
Figure \ref{fig:length_error} indicates that the distance error is between $1-3\text{mm}$ for the level-sets chosen
in our numerical example. These levelsets were chosen to replicate the current treatment radius of
about 2cm.

\section{Discussion} \label{sec:discussion}
%\paragraph*{Parameters and Datasets}~\newline
We presented a stationalisation approach for the estimation of the glioblastoma invasion extent.
The stationalisation approach partially addresses the problems
of para\-metri\-zation and data availability. 
The stationary simulations do not depend on the \textit{complete} 
knowledge of the initial condition to produce reasonable tumor invasion estimates.
The thresholded information provided by the medical images 
might be fully utilized with the limited information it provides.
The stationalisation only requires 
datasets from one point in time, i.e. one DTI scan and a medical 
segmentation of the tumorous region. This is the data that is routinely 
gathered in medical practice, as it is used for 
planning the radiation therapy and resection. 
Stationary simulations, as presented here, may provide an additional tool 
in this regard without altering the imaging practices.

The problem of quantifying model parameters is also partly alleviated by the 
fact that in the stationary formulation the solution depends merely on the 
strength of the parameters with respect to each other, and not their absolute values. 
It may be easier to experimentally determine ratio values between the three
important parameterized terms: $\mathbf{D}_t, v$ and $\rho$ than the full set 
of parameters needed for a forward simulation.
Forward models which fit into the form of equation \eqref{eq:myopic_diff_react} 
may only produce reliable results if the correct initial condition $g(\mathbf{x})$ is 
known and the parameters are determined to a reasonable degree of accuracy. 
In the medical setting the only information at 
hand is the medical imaging at time of diagnosis. 
For the forward models to produce reasonable estimates on the tumor 
invasion profile we would firstly need information on the location and time of the carciogenesis, 
secondly the information on the non-degraded diffusive 
properties of the tissue surrounding it and 
lastly the correct parametrisation on a per-patient basis. 
We want to emphasise that the temporal 
dynamics of the tumor growth, 
although scientifically interesting, are not necessarily 
relevant in the medical treatment planning. 
The information needed for treatment with the current techniques is an 
accurate description of the tumor density field \textit{at time of diagnosis}.

%\paragraph*{Applicability}~\newline
The derivation of the stationalisation term for the one dimensional 
Fisher equation is not strictly transferable to tumor models which 
incorporate medical data or have additional advection terms, 
but the results from section \ref{subsec:3D_real_data} indicate 
that they may still produce reasonable predictions. 
If the underlying diffusion matrix field included strong inhomogeneities 
inducing strong advective terms, 
the procedure might lose its validity, however an investigation 
of the given 3D DTI dataset shows that the peclet-number relating
the drift and diffusion strength as given in \eqref{eq:stationary_fisher_with_D}, $\tau =
\frac{|v~L|}{|D|}$, stays mostly below 0.3 when $\mathbf{D}_t$ is derived from the
DTI data by simple scaling \eqref{eq:D_t}.
Most tumor models show travelling wave characteristics, 
where the main physical effects include diffusion and nonlinear growth.
We recommend close inspection of both the gradient distributions
and the peclet numbers if the stationalised model
should be extended. The stationalisation procedure should retain 
its applicability as long as the gradient distribution retains 
its underlying characteristic, and the distance between
medical segmentation and the border induced by the level-sets
is not chosen too large. In the case that the presented approximation
fails, it may also be possible to
introduce more elaborate numerical ways to fit the necessary
penalty term locally.

We compared the level-sets of the forward simulation with those of the 
stationary simulation and found characteristic distances between them 
of about $L_B \approx$ 1.0-3.0mm (Fig. \ref{fig:length_error}).
Compared to a fixed-size radius of 2cm around the visible tumor 
\citep{ChangAkyu2007}, these errors seem justifiable. It
is of course possible to find an optimal penalty parameter $v$
for a given set of forward model parameters.
A medical practitioner might choose the actual level-set value to 
replicate the current practice of treating a 2 cm radius around the bulk tumor and 
then use qualitative information in the form of locally recommending 
an extension or retraction of the treatment radius. 
The stationary model will correctly capture the effect of the material 
properties, as presented in Figure \ref{fig:butterfly}. 
Where the tissue is more diffusive, the level-set on $u_s$ 
will overextend the 2 cm radius and where the diffusivity is small, 
more brain tissue might be left untreated. 
If the underlying tumor model should be extended, 
all effects increasing- or diminishing the spread of glia cells will 
be reflected accordingly in the stationalised solution.

Should time-series datasets become available, then the stationalisation may
 be used to estimate the initial condition
$g(\mathbf{x})$ for a forward simulation from the first dataset.
An initial condition calculated in this way should be closer to
a presumably smooth real density profile than using a stepped profile
with steep gradients.
 
%\paragraph*{Performance}~\newline
Naturally the computation of solutions to the stationary 
problem take less time than a full forward simulation. 
While compute servers are usually available in academic institutions, 
the possibility to calculate the results on a regular consumer pc with
only short computation times
is important in order to transfer such methods into clinical application. 
The stationary simulation allows for the computation of sets of 
solutions for varying parameters within a short timeframe. 
We present exemplary runtimes for the camino dataset on recent 
hardware for the two cases in Table \ref{tab:runtimes}.

\begin{table}[h]
\begin{center}
\begin{tabular}{lll}
\hline\noalign{\smallskip}
\textbf{simulation type} & \textbf{hardware} & \textbf{runtime} \\
\noalign{\smallskip}\hline\noalign{\smallskip}
2D, 90 days	 & Intel i5-7200U  ( 4x2.50GHz )& 42 sec\\
2D, stationary & Intel i5-7200U  ( 4x2.50GHz )& 0.4 sec\\
\noalign{\smallskip}
3D, 90 days	& Intel i5-7200U  ( 4x2.50GHz )& 1:31h\\
3D, stationary& Intel i5-7200U  ( 4x2.50GHz )& 76sec\\
\noalign{\smallskip}
3D, 90 days	& AMD EPYC 7501 (32x 2.0GHz)& 18min\\
3D, stationary& AMD EPYC 7501 (32x 2.0GHz)& 9 sec\\
\noalign{\smallskip}\hline
\end{tabular}
\caption{Runtimes for the testcases in two- and three dimensions on different hardware.}
\label{tab:runtimes}
\end{center}
\end{table}

\subsection{Outlook}
There might be further improvements to the stationalisaton approach. 
Altering the stationalisation term, which currently assumes a globally constant wave-propagation speed, 
to be sensitive to the local material properties might further improve the results. 
The proportionality of the wave speed ($v\geq 2~\sqrt{D \rho}$) in 
the one dimensional Fisher-KPP equation may be an indicator for how a localized penalty parameter could be improved. 
Instead of choosing a constant $v$ globally, it might be 
possible to set a penalty factor linearly combined with local information. 
Thereby incorporating the local increases and decreases in diffusivity into the stationalisation.

In section \ref{subsec:3D_real_data} we used a real dataset, but a comparatively primitive tumor model. 
It should be possible to extend the stationalisation procedure to 
incorporate additional effects like chemo- or haptotaxis as long as the dynamic 
of producing travelling wave solutions of sigmoidal shape is not altered by the additions. 
In the example in section \ref{subsec:3D_real_data}, 
we used a level-set on the forward simulation as the internal Dirichlet constraints for the stationalisation. 
In a medical setting one would directly use the medical segmentation 
from the DTI/MRI modalities. There are promising advances in generating 
tumor segmentations in an automated fashion, e.g. BraTumIA\footnote{\url{https://www.nitrc.org/projects/bratumia}} \citep{BraTumIA}.
Automating the process of the segmentation opens up the possibility 
to use a fully automated process to advise the treatment planning in real patients.

We showed how well a stationalised formulation would perform compared 
to a forward simulation if all the necessary information were present.
The fact that time-series datasets are largely unavailable and therefore
no parametrizations can be derived from them,
 makes direct comparisons between existing tumor models difficult.
It is, however possible to compare the stationalized versions of
existing models with only the datasets from the time of diagnosis.
One could then compare these levelsets to the clinical target volume (CTV)
 regularly produced in medical practice. This offers an attractive
 approach to perform a model comparison for a wide range of tumor
 models proposed in the literature.

\begin{acknowledgements}~\\
The authors would like to express gratitude towards the BMBF (Bundesministerium f\"ur Bildung und Forschung) for 
funding the GlioMath project (funding code: 05M16PMA),
and all collaborating partners within.
\end{acknowledgements}

\bibliography{./gliomatLitDatabase}
\bibliographystyle{spbasic}      % basic style, author-year citations
\end{document}